\documentclass[conference]{IEEEtran}
\IEEEoverridecommandlockouts
\usepackage{cite}
\usepackage{amsmath,amssymb,amsfonts}
\usepackage{algorithmic}
\usepackage{graphicx}
\usepackage{textcomp}
\usepackage{xcolor}

\usepackage{url}
\usepackage{extra_styles}
\usepackage{booktabs}
\usepackage{algorithm}
\usepackage{algorithmic}
\usepackage[breaklinks, colorlinks, pagebackref]{hyperref} 
\usepackage[subtle,mathspacing=normal]{savetrees}

\def\BibTeX{{\rm B\kern-.05em{\sc i\kern-.025em b}\kern-.08em
    T\kern-.1667em\lower.7ex\hbox{E}\kern-.125emX}}
\begin{document}

\title{VoiceDiT: Dual-Condition Diffusion Transformer \\for Environment-Aware Speech Synthesis
}

\author{\begin{tabular}{c}
\textit{Jaemin Jung$^{*}$, Junseok Ahn$^{*}$, Chaeyoung Jung, Tan Dat Nguyen, Youngjoon Jang, Joon Son Chung} \\
Korea Advanced Institute of Science and Technology, South Korea\\
\{jjm5811, junseok.ahn, codud9914, tandat.kaist, wgs01088, joonson\}@kaist.ac.kr
\end{tabular}
\thanks{
$^*$These authors contributed equally to this work.
This work was supported by the National Research Foundation of Korea (NRF) grant funded by the Korea government (MSIT) (No. RS-2023-00222383).}
\cr
\vspace{-10pt}
}

\maketitle

\begin{abstract}
We present VoiceDiT, a multi-modal generative model for producing environment-aware speech and audio from text and visual prompts. While aligning speech with text is crucial for intelligible speech, achieving this alignment in noisy conditions remains a significant and underexplored challenge in the field. To address this, we present a novel audio generation pipeline named VoiceDiT. This pipeline includes three key components: (1) the creation of a large-scale synthetic speech dataset for pre-training and a refined real-world speech dataset for fine-tuning, (2) the Dual-DiT, a model designed to efficiently preserve aligned speech information while accurately reflecting environmental conditions, and (3) a diffusion-based Image-to-Audio Translator that allows the model to bridge the gap between audio and image, facilitating the generation of environmental sound that aligns with the multi-modal prompts.
Extensive experimental results demonstrate that VoiceDiT outperforms previous models on real-world datasets, showcasing significant improvements in both audio quality and modality integration. 
Synthesized samples are available on our demo page: \url{https://mm.kaist.ac.kr/projects/voicedit/}
\end{abstract}


\section{Introduction}
The demand for generating realistic audio, including sound effects, music, and speech, has rapidly increased across various industries, such as film and gaming. In response, the fields of Text-to-Audio (TTA)~\cite{yang2023diffsound, kreuk2022audiogen, huang2023make, ghosal2023text, evans2024fast,  liu2024audiolcm, xue2024auffusion} and Text-to-Speech (TTS)~\cite{popov2021grad, kim2020glow, kim22guidedtts, huang2023prosody, kim2023crossspeech, nguyen2024fregrad, Ju24naturalspeech, lee2024ditto} have garnered significant attention for their ability to generate natural sounds and speech from text prompts. 
Recently, diffusion models, which have been extensively studied in the field of image generation~\cite{rombach2022high, song2020score, ho2020denoising}, have emerged as a transformative approach to audio synthesis. Grad-TTS~\cite{popov2021grad} leverages diffusion processes to produce high-quality speech from text, and AudioLDM~\cite{liu2023audioldm} enables the versatile generation of environmental sounds and music from text prompts.

While these models perform well at their respective tasks, audio generation often requires the simultaneous generation of speech and environmental sounds. For example, to enhance realism in extended reality (XR) applications, the character's voice must blend naturally with environmental sounds and reflect the acoustics of the space. Recent studies~\cite{lee2024voiceldm, vyas2023audiobox, liu2024audioldm, yang2023uniaudio} have explored specialized model architectures capable of jointly generating both audio and speech. 
VoiceLDM~\cite{lee2024voiceldm} conditions the U-Net backbone of AudioLDM with speech transcriptions through cross-attention, enabling the simultaneous generation of both speech and environmental sounds. 
It uses the contrastive language-audio pre-training (CLAP)~\cite{wu2023large} model to train with sound clips instead of text annotations, addressing the issue of data scarcity.
However, VoiceLDM often produces repetitive and mumbled speech due to a lack of temporal alignment between speech and text.
More recently, AudioBox~\cite{vyas2023audiobox} introduces a unified model based on flow-matching, capable of generating various audio modalities. 
AudioBox is trained on data annotated by a large language model or human experts, encompassing diverse speech attributes such as age, gender, and accent. However, as the labels primarily focus on speaker style rather than environmental context, AudioBox's ability to generate a wide range of environmental sounds remains limited.

In this paper, we present VoiceDiT, a novel audio generation pipeline that not only produces natural speech but also allows for conditioning on the acoustic environment. The proposed pipeline features three key components: (1) data synthesis and refinement for limited training data, (2) a transformer-based architecture designed for multi-task performance, and (3) an image-to-audio translator~(I2A-Translator) to enhance flexibility in handling diverse inputs.

Current efforts to integrate TTA and TTS systems are constrained by the lack of large speech datasets that are both accurately transcribed and reflective of the diverse audio conditions encountered in real-world scenarios~\cite{ zen2019libritts, ardila2019common, nagrani2017voxceleb, chung2018voxceleb2, kwak2024voxmm}.
To address this issue, we propose a practical strategy for data acquisition by first constructing a large-scale synthetic dataset, where environmental noise and reverberation are added to clean speech for pre-training the model. 
Subsequently, we refine the real-world speech dataset\cite{lee2024voiceldm} for fine-tuning, which helps bridge the domain gap between synthetic and real-world data.

We then introduce a Dual-condition Diffusion Transformer~(Dual-DiT), designed to generate environment-aware speech by incorporating two distinct conditions: one for speech and another for environmental sound. 
For stable and efficient speech generation, we develop a TTS module that meticulously aligns text with speech using alignment information~\cite{kim2020glow}, along with a Latent Mapper that compresses long text conditions into a latent space.
Additionally, to ensure the model generates sounds suitable for the given environmental conditions, we integrate a cross-attention module into the Dual-DiT blocks. This module injects environmental features, enabling the model to generate sounds that align with the conditions.

Finally, to broaden our model’s applicability, we introduce a diffusion-based I2A-Translator that converts image embeddings into audio embeddings. Unlike text, images provide a more intuitive representation of environments, especially for complex or abstract scenarios. Through the I2A-Translator, our model can generate diverse and nuanced audio conditioned on both text and images.

Our comprehensive experimental results demonstrate that VoiceDiT excels at generating environment-aware speech that aligns closely with user prompts, achieving state-of-the-art performance across both qualitative and quantitative metrics. This success underscores the potential of VoiceDiT for a wide range of applications.

\section{Method}
\subsection{Data Preparation}

\newpara{Pre-training dataset.}
To train the TTS components that generate intelligible speech, it is essential to have a speech dataset aligned with corresponding transcripts. However, large-scale real-world transcribed speech datasets are currently scarce. To address this shortage of data, we construct a large-scale synthetic dataset by adding various noises to clean speech.
For the noise data, we utilize WavCaps~\cite{mei2024wavcaps}, a sound event dataset with audio captions. Since WavCaps contains a notable amount of speech data alongside environmental sounds, we filter out speech-related samples, retaining 340K environmental sound samples.
We then mix clean speech from LibriTTS-R~\cite{koizumi2023libritts} with the noise data, using a signal-to-noise ratio value randomly selected from a uniform distribution within the range of 2 to 10. Furthermore, to simulate various environmental conditions, we apply Room Impulse Response filters with a certain probability.

\newpara{Fine-tuning dataset.}
To bridge the domain gap between synthetic and real-world data, we further fine-tune our model on the in-the-wild speech dataset, AudioSet-speech~\cite{lee2024voiceldm}. This is a speech subset of AudioSet transcribed with ASR models.
Despite a carefully designed ASR process for accurate transcription, as detailed in~\cite{lee2024voiceldm}, AudioSet-speech still contains many inaccurate transcriptions.
Additionally, since the audio length is fixed at 10 seconds, non-speech segments are present before and after the speech segments within the audio sample.
Such misaligned data hinders the alignment of speech and text during the training of TTS model.
To address these issues, we implement a two-step preprocessing approach~\cite{jung2023metric}.
First, we compute the word error rate (WER) on AudioSet-speech dataset using the Whisper \textit{large-v3} model~\cite{radford2023robust}, filtering out samples where the WER exceeds 20\%.
This process reduces the original 597K speech data to a refined dataset of 400K, with most of the discarded data consisting of songs or audio featuring multiple speakers.
Next, we use a forced aligner to align the transcript of spoken words with the corresponding audio recording. Then, we truncate non-speech segments from the beginning and end of the speech segment, ensuring proper alignment between the speech and transcripts.


\begin{figure}[t]
\label{main_fig}
  \centering
  \includegraphics[width=0.95\linewidth]{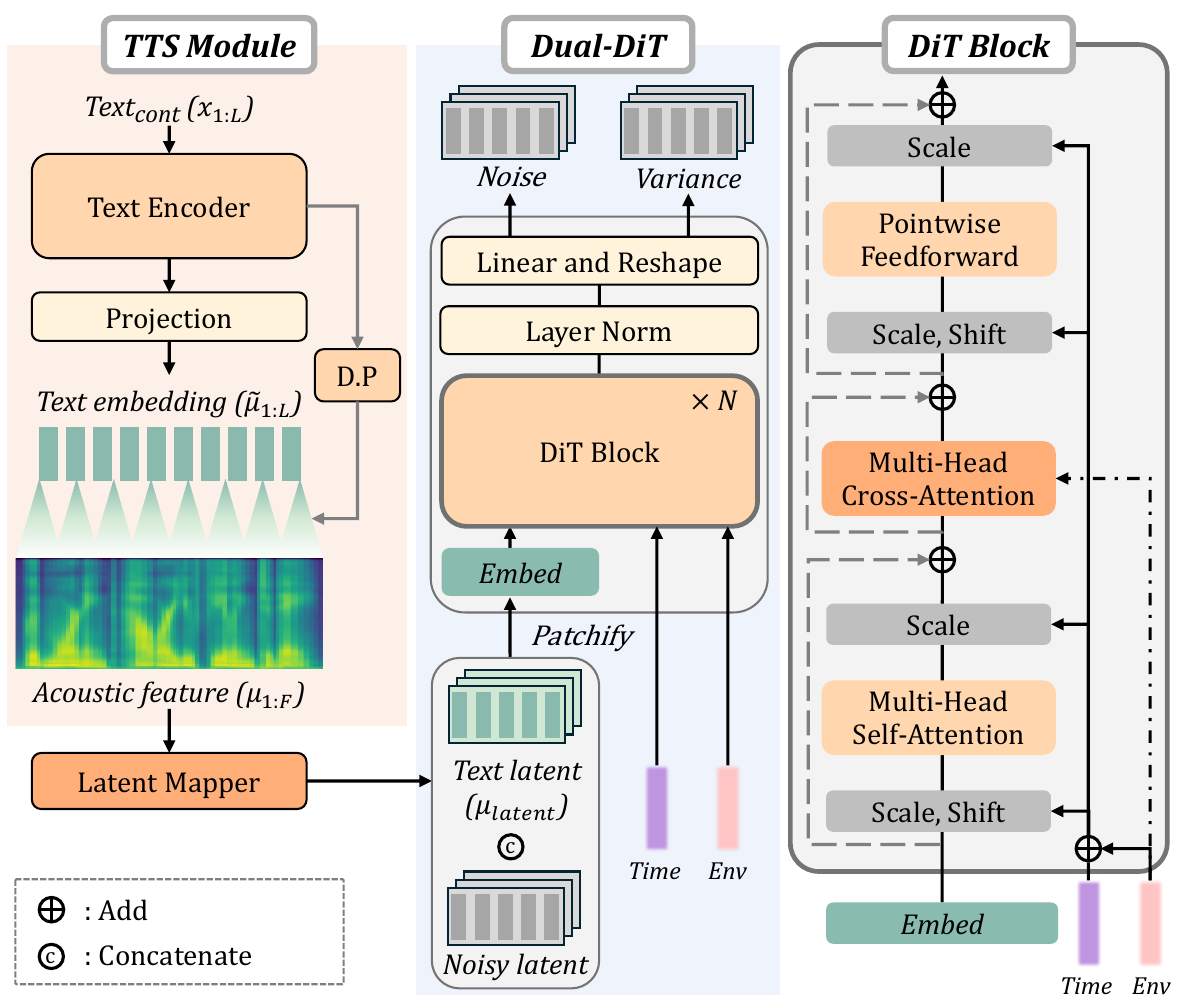}
  \vspace{-3mm}
  \caption{
  \textbf{Model architecture of VoiceDiT.} VoiceDiT consists of a TTS module and a Dual-DiT model. 
  A cross-attention module is integrated into each DiT block to inject environmental conditions. ``D.P'' stands for Duration Predictor.}
  \label{fig:framework}
  \vspace{-5mm}
\end{figure}

\subsection{Model Architecture}
VoiceDiT is a latent diffusion model with two components: the TTS module, which generates text-aligned acoustic features, and Dual-condition Diffusion Transformer (Dual-DiT), a decoder that synthesizes environment-aware speech from two input conditions. An overview of the architecture is provided in \Fref{fig:framework}.

\newpara{TTS module.}
The TTS module consists of a text encoder and a duration predictor, both adopted from Glow-TTS~\cite{kim2020glow}. This module is responsible for extracting time-aligned linguistic information to generate the mel-spectrogram \(y_{1:F}\) from the input text sequence \(x_{1:L}\), where \(F\) and \(L\) represent the lengths of the respective sequences. 
First, the text encoder extracts linguistic information \(\tilde{\mu}_{1:L}\) from the input text sequence \(x_{1:L}\). Then, the upsampled encoded text $Up(\tilde{\mu}_{1:L}) = \tilde{\mu}_{1:L, A^*}$ is obtained according to the best alignment \(A^*\) between \(\tilde{\mu}_{1:L}\) and the mel-spectrogram \(y_{1:F}\). During training, the alignment \(A^*\) is determined using MAS. 
The text encoder is trained to maximize the probability $p(y_{1:F}; \tilde{\mu}_{1:L, A^*}, \boldsymbol{I})$ by minimizing the encoder loss ($L_{enc}$).
Furthermore, the optimal alignment is used to extract ground truth duration $d$, which supervises the training process of duration predictor by loss function $\mathcal{L}_{dp} = \Vert \log(d) - \log(\hat{d})\Vert_2^2$, where $\hat{d}$ is predicted duration conditioned on $\tilde{\mu}_{1:L}$. 
During inference, the duration predictor replaces MAS to provide alignment since only text is provided at this stage.

\newpara{Dual-condition diffusion transformer.}
We present the Dual-DiT, a model designed to fuse two distinct conditions to generate natural and coherent speech. We begin by adopting a Transformer-based architecture~\cite{peebles2023scalable} as our base model, utilizing its multi-head attention-based fusion mechanism, which is effective at modeling long-range dependencies and facilitating efficient modality fusion~\cite{chen2023pixartalpha}.

Subsequently, we explore two methods for incorporating the acoustic feature into the DiT model.
When long acoustic features are integrated via cross-attention, the computational complexity scales quadratically with the sequence length~\cite{beltagy2020longformer}.
Alternatively, we concatenate the acoustic features with the noisy input in the mel-spectrogram space before passing them through the DiT model. This method yields more stable training and produces natural-sounding speech.
However, training the diffusion model in the high-resolution mel-spectrogram space remains computationally demanding.
To address this, we propose a Latent Mapper, consisting of two 2D convolutional layers, to map the long acoustic features into the latent space of the DiT.
Specifically, the acoustic feature \(\mu_{1:F}\) of size $T\times F$ is mapped into a latent representation \(\mu_{latent}\) with a resolution of $8 \times T/4 \times F/4$, aligning with the resolution of the noisy latent.
By employing the Latent Mapper, we enable the diffusion model to be trained in the latent space, significantly reducing the computational cost by 94\% compared to pixel-space-based DiT models.

To enable the model to generate sounds appropriate for the given environment conditions, we design the DiT block by incorporating a cross-attention module~\cite{vaswani2017attention}. In our implementation, environmental conditions are extracted using the Contrastive Language-Audio Pre-training (CLAP) audio encoder. 
The original DiT processes this additional conditional information through adaptive layer norm (adaLN) modules where the scale and shift parameters are regressed from the sum of the time embedding and condition label.
However, due to adaLN's reliance on affine transformations, there is a potential risk of losing detailed conditioning information~\cite{perez2018film}. To address this issue, we introduce a cross-attention module between the self-attention and feed-forward layers. This design choice ensures that CLAP embeddings are directly injected into the DiT model via cross-attention, thereby facilitating a more precise representation of detailed environmental sounds.

\subsection{I2A-Translator}
To broaden our model’s applicability, we introduce a diffusion-based I2A-Translator, an expert network that converts image latent into audio latent.
Following V2A-Mapper~\cite{wang2024v2a}, we train a Transformer network \(f_{i2a}\) to predict the CLAP audio embedding \(z_0\) from a Contrastive Language-Image Pre-Training (CLIP)~\cite{radford2021learning} image embedding \(y\). The training objective is formulated as \(L_{i2a} = \mathbb{E}_{t \sim [1,T]} \left[ \Vert z_0 - f_{i2a} (t, z_t, y) \Vert \right]\).
In this setup, the input of the Transformer comprises timestep embedding \(t\), noisy audio embedding \(z_t\), image embedding \(y\), and a learnable token. The output from the network at the learnable token position is considered to be the recovered audio embedding during inference. This component enables the model to accept visual prompts, significantly expanding its range of applications and enhancing its versatility and adaptability to various practical scenarios. 


\subsection{Training}
The training process of VoiceDiT mostly follows the framework of VoiceLDM~\cite{lee2024voiceldm}, which takes two conditions as input. First, an input audio is encoded into a latent representation \(z_0\) by a pre-trained variational autoencoder (VAE) encoder. A noisy latent of \(z_t\) at a specific timestep $t$ is obtained through the forward diffusion process by adding noise to \(z_0\) according to a predefined noise schedule.

The TTS module and Latent Mapper encode the content prompt \(text_{cont}\) to form the text latent \(\mu_{latent}\), which serves as the content condition \(\mathbf{c}_{cont}\). This text latent is concatenated with the noisy latent \(z_t\) and fed into the DiT model.
Meanwhile, the CLAP audio encoder directly processes the input audio to derive the environment condition \(\mathbf{c}_{env}\), eliminating the need for a manually annotated prompt \(text_{env}\)~\cite{liu2023audioldm}.
Finally, the model is trained to predict the added noise $\boldsymbol{\epsilon}$ using the re-weighted training objective as follows:

\vspace{-2mm}
\begin{equation}
\mathcal{L}_{diff} = \|\boldsymbol{\epsilon} - \boldsymbol{\epsilon}_\theta(z_t, t, \mathbf{c}_{env}, \mathbf{c}_{cont})\|_2^2. 
\end{equation}

The parameters of the Dual-DiT, TTS module, and Latent Mapper are jointly optimized, while the other components remain frozen.

\subsection{Inference}

Our Dual-DiT samples new latent conditioned on \(\mathbf{c}_{env}\) and \(\mathbf{c}_{cont}\).
The condition \(\mathbf{c}_{cont}\) is generated from a content text prompt following the same procedure as in training, while \(\mathbf{c}_{\text{env}}\) is constructed by extracting CLAP embedding from each modality input. The CLAP text encoder is employed for text inputs, whereas the CLIP image encoder is used for image inputs, with the embeddings mapped to the CLAP embedding space via the I2A-Translator. To enhance the controllability and flexibility of each condition, we employ the dual classifier-free guidance~\cite{lee2024voiceldm}.
Specifically, given the two conditions \(\mathbf{c}_{\text{env}}\) and \(\mathbf{c}_{\text{cont}}\), the diffusion score \(\boldsymbol{\epsilon}_\theta\) is adjusted as follows:
\vspace{-2mm}
\begin{align}
\nonumber\hat{\boldsymbol{\epsilon}}_\theta(z_t, &\mathbf{c}_{env}, \mathbf{c}_{cont}) = \boldsymbol{\epsilon}_\theta(z_t, \mathbf{c}_{env}, \mathbf{c}_{cont})\\\nonumber
&+ w_{env}\Big(\boldsymbol{\epsilon}_\theta(z_t, \mathbf{c}_{env}, \emptyset_{cont})-\boldsymbol{\epsilon}_\theta(z_t, \emptyset_{env}, \emptyset_{cont})\Big)\\
&+ w_{cont}\Big(\boldsymbol{\epsilon}_\theta(z_t, \emptyset_{env}, \mathbf{c}_{cont})-\boldsymbol{\epsilon}_\theta(z_t, \emptyset_{env}, \emptyset_{cont})\Big),
\end{align}
where \(w_{env}\) and \(w_{cont}\) are the guidance scale for environment and content conditions, respectively, \(\emptyset\) indicates the null condition. 
After generating a new latent vector from Dual-DiT, it is converted to mel-spectrogram space using the VAE decoder. Finally, the pre-trained HiFi-GAN vocoder~\cite{kong2020hifi} transforms the mel-spectrogram into the desired audio.

\section{Experiments}
\begin{table}[t]
\centering
\caption{
Performance evaluation on the AC-Filtered.  
MOS results are presented with a 95$\%$ confidence interval.
$\uparrow$ denotes higher is better, $\downarrow$ denotes lower is better.}
\vspace{-2mm}
\resizebox{0.95\linewidth}{!}{\begin{tabular}{l|ccc|ccc}
\toprule
Model & FAD$\downarrow$      & CLAP$\uparrow$ & WER(\%)$\downarrow$ & Nat.$\uparrow$ & Intel.$\uparrow$ & REL $\uparrow$ \\ \midrule
GT & -            & 0.40     & 17.47    & 4.24 {\scriptsize $\pm 0.10$}    &  4.08 {\scriptsize $\pm 0.11$}    & 4.26 {\scriptsize $\pm 0.09$}   \\ \midrule
VoiceLDM & 5.56  &   0.21   &   10.39  & 2.94 {\scriptsize $\pm 0.11$}   & 3.35 {\scriptsize $\pm 0.12$}     &  3.24 {\scriptsize $\pm 0.11$}  \\
VoiceDiT & \bf{4.60}         & \bf{0.22}     & \bf{7.09}    & \bf{3.41} {\scriptsize $\pm 0.10$}   & \bf{4.32} {\scriptsize $\pm 0.08$}     & \bf{3.86} {\scriptsize $\pm 0.09$}  \\ 
\bottomrule
\end{tabular}}
\label{table:main}
\vspace{-5mm}
\end{table}

\subsection{Experimental Setup}
\newpara{Data description.} 
To pre-train the VoiceDiT, we use the processed WavCaps, which contains 340K non-speech data, and LibriTTS-R, a multi-speaker corpus with 585 hours of speech data.
For fine-tuning, we utilize AudioSet-speech, a real-world speech dataset comprising 597K samples, refined to 400K for training. To train the I2A-Translator, we use the VGGSound dataset, which contains 165K 10-second video clips. CLIP image embeddings are extracted from 10 frames per video, averaged along the time axis for input.

\newpara{Implementation details.}
The \texttt{DiT\_L/2} model, consisting of 24 DiT blocks with a hidden dimension of 1,024, serves as the backbone network. We adopt a VAE to encode the mel-spectrogram into a latent representation, which is then divided into patches of size 2 for the DiT input. For the speaker conditioning, a 192-dimensional speaker embedding is extracted using \texttt{WavLM-ECAPA}~\cite{chen2022wavlm} model from ESPnet-SPK toolkit~\cite{jung2024espnet}. The entire model is pre-trained for 100K steps using synthetic data exclusively and then fine-tuned for 20K steps with additional data from AudioSet-speech.
During pre-training, duration loss ($L_{dp}$), encoder loss ($L_{enc}$), and diffusion loss ($L_{diff}$) are employed to achieve text-to-speech alignment. After this phase, the converged TTS module is kept frozen, and only the diffusion loss is applied.
We use AdamW optimizer~\cite{loshchilov2017decoupled} with a constant learning rate of $ 10^{-4}$ for pre-training and halved it at fine-tuning. Our model is trained on 8 NVIDIA A6000 GPUs with a batch size of 16. In inference, we empirically set $w_{desc} = 5$, $w_{cont} = 5$ for dual classifier-free guidance.

\newpara{Evaluation metrics.}
For objective evaluation, we utilize Frechet Audio Distance (FAD)~\cite{kilgour2018fr} and Kullback-Leibler (KL) divergence to assess audio quality, and the CLAP score to measure the correspondence between the text description and the generated audio. Additionally, to evaluate speech intelligibility, we calculate the word error rate (WER) of the synthesized speech against the ground truth using the Whisper \textit{Large-v3} model. For subjective evaluation, we assess three main aspects of speech synthesis with the environment: Mean Opinion Score (MOS) of Speech Naturalness (Nat.), MOS of Speech Intelligibility (Intel.), and Relevance to text description (REL)~\cite{liu2023audioldm}. 
During these evaluations, 20 participants rate the naturalness, intelligibility, and the match between environmental sounds and their corresponding descriptive text on a scale of 1 to 5.
The subjective evaluation is conducted on 40 randomly selected samples per model using a unified set of prompts.

\section{Results}

\begin{table}
\centering
\vspace{-1mm}
\caption{The comparison between conditioning methods on AC-Filtered.
{\bf Text}: Text conditioning method,
{\bf Env}: Environment conditioning method,
{\bf CAT}: Concatenation,
{\bf CA}: Cross-attention.}
\vspace{-2mm}
\resizebox{0.90\linewidth}{!}{\begin{tabular}{ccc|cccc}
\toprule
\multicolumn{1}{c}{Model} & Text   & Env & FAD$\downarrow$ & KL$\downarrow$ & CLAP$\uparrow$ & WER$(\%)\downarrow$ \\ \midrule 
\multicolumn{1}{c}{U-Net} & CAT    & add & 7.88 & 2.22 & 0.13  & 9.28 \\ \cmidrule{1-7}
\multirow{3}{*}{Dual-DiT} & CA     & adaLN & 4.23 & 1.44 & 0.24 & 94.80 \\
                          & CAT     & adaLN    &  4.89   &  1.51  &   0.23   &  10.33   \\
                          & CAT     & adaLN+CA    &  4.61   &  1.49  &   0.22   &  7.09   \\
\bottomrule
\end{tabular}}
\label{table:condition}
\vspace{-5mm}
\end{table}

\subsection{Comparison with State-of-the-arts}
We compare our VoiceDiT model with VoiceLDM on AC-filtered~\cite{lee2024voiceldm}, which is created by transcribing the speech data from the AudioCaps test set. As shown in \Tref{table:main}, our model outperforms VoiceLDM across all metrics, with significant improvements of 32.9\% in WER and 28.96\% in intelligibility. VoiceLDM does not ensure alignment between speech and text, leading to low-quality outputs including repetitive speech. In contrast, by explicitly aligning speech and text through our TTS module, VoiceDiT generates more intelligible speech. Remarkably, it even achieves a higher intelligibility score compared to the ground truth.

\subsection{Ablation Study} 
\newpara{Model architecture.}
To assess the performance differences between the Dual-DiT and the U-Net architecture, we train the U-Net with all settings identical except for the model structure.
The U-Net architecture follows the setup from \cite{liu2023audioldm}, with encoder block channel dimensions of \([c_u, 2c_u, 3c_u, 5c_u]\), where \(c_u\) is the base channel number, set to 256 in our experiments. 
The U-Net model has a total of 467M parameters, while the Dual-DiT model contains 457M parameters.
Comparing the first and third rows of \Tref{table:condition}, we observe that the Dual-DiT model outperforms the U-Net architecture across all objective metrics except WER. 
This indicates that DiT leverages the power of the multi-head attention mechanism to capture long-range dependencies and spatial-temporal representations, resulting in temporally consistent and high-quality audio outputs.

\newpara{Text conditioning methods.}
We evaluate two approaches for conditioning the text latent from the TTS Module on the DiT model. Initially, we inject the text latent directly into the cross-attention module of the Dual-DiT block. Subsequently, we concatenate the text latent with the noisy latent along the channel dimension, feeding this combined input into the Dual-DiT.
As shown in the second and third rows of \Tref{table:condition}, the cross-attention method generates completely unintelligible speech, resulting in a WER of 94.80\%. In contrast, when the text latent is concatenated into the DiT input, the model produces sufficiently intelligible speech after the same number of training steps, achieving a WER of 10.33\%.
These results demonstrate that the proposed method effectively preserves the sequence information of speech content, facilitates faster learning compared to the 3M training steps required by previous work~\cite{lee2024voiceldm}, and significantly aids in generating intelligible speech.
Note that the environmental conditions are controlled solely through the adaLN module.

\newpara{Environmental conditioning methods.}
We perform an ablation study on the cross-attention module within the DiT block to evaluate its impact on generating environment-aware speech.
As detailed in the third and fourth rows of \Tref{table:condition}, the cross-attention module leads to improved performance across the FAD, KL divergence, and WER metrics.
This enhancement is primarily due to the cross-attention module's efficacy in integrating CLAP embeddings. The cross-attention module captures comprehensive environmental details, compensating for the limitations of the adaLN, which tends to lose intricate details due to its reliance on simple affine transformations.

\subsection{X-to-Audio Capabilities}
First, we evaluate the zero-shot TTA generation capabilities of VoiceDiT on the AudioCap test set, applying dual classifier guidance with $w_{\text{desc}}=9$ and $w_{\text{cont}}=1$. 
As summarized in \Tref{table:modality}, VoiceDiT outperforms VoiceLDM across all evaluation metrics, demonstrating significantly superior performance. Moreover, despite not being explicitly trained for TTA, our approach achieves better FAD and KL divergence scores compared to AudioLDM~\cite{liu2023audioldm}.
Next, we assess the image-to-audio generation capabilities of VoiceDiT using the VGGSound test set.
Utilizing the I2A-Translator for image-based prompts, VoiceDiT achieves a comparable KL divergence score and a superior FAD score relative to Im2Wav~\cite{sheffer2023hear}, the state-of-the-art image-to-audio generation model. These findings demonstrate the versatility of our model in generating high-quality audio from diverse multi-modal inputs, including text and images.

\section{Conclusion}

\begin{table}
\centering
\vspace{-1mm}
\caption{Performance comparison of text-to-audio capabilities on the AudioCaps test set and image-to-audio capabilities on the VGGSound test set. The results for $\dag$ are as reported in ~\cite{iashin2021taming}, \cite{sheffer2023hear}.}
\vspace{-2mm}
\resizebox{0.80\linewidth}{!}{\begin{tabular}{ccc|cccc}
\toprule
\multicolumn{1}{c}{Modality}    & Model     &Params & FAD$\downarrow$ & KL$\downarrow$ & CLAP$\uparrow$ \\ \midrule 
\multicolumn{1}{c}{-} & GT & - & -    & - & 0.52  \\ \midrule
\multirow{3}{*}{Text} & AudioLDM~\cite{liu2023audioldm} & 541M & 4.27 & 2.01 & 0.42  \\
                          & VoiceLDM~\cite{lee2024voiceldm}    & 508M & 10.91   & 2.95  &   0.29      \\
                          & VoiceDiT (\textbf{Ours})    & 565M & \textbf{3.55}   & \textbf{1.87}  &  \textbf{0.45}     \\ \midrule
\multirow{3}{*}{Image}& SpecVQGAN~\cite{iashin2021taming}$^{\dag}$ &379M   & 6.64  & 3.10 & -  \\  
                      & Im2Wav~\cite{sheffer2023hear}$^{\dag}$    & 360M  & 6.41 & \textbf{2.54} & -  \\
                      & VoiceDiT (\textbf{Ours})   & 565M & \textbf{3.02} & 2.73 & -  \\
\bottomrule
\end{tabular}}
\label{table:modality}
\vspace{-5mm}
\end{table}

In this paper, we propose VoiceDiT, a multi-modal generative model that generates environment-aware speech from text, audio, and visual prompts. The Dual-DiT model effectively preserves speech alignment while adapting to environmental conditions, and the I2A-Translator enhances the model’s versatility by incorporating visual inputs. Experimental results demonstrate that VoiceDiT significantly outperforms existing methods in both audio quality and multi-modality integration. Our research facilitates adaptive speech generation for various acoustic environments, offering promising applications in filmmaking and audiobook production.

\clearpage

\bibliographystyle{IEEEbib}
\bibliography{refs}

\begin{thebibliography}{10}

\bibitem{yang2023diffsound}
Dongchao Yang, Jianwei Yu, Helin Wang, Wen Wang, Chao Weng, Yuexian Zou, and Dong Yu,
\newblock ``Diffsound: Discrete diffusion model for text-to-sound generation,''
\newblock {\em IEEE/ACM Trans. on Audio, Speech, and Language Processing}, vol. 31, pp. 1720--1733, 2023.

\bibitem{kreuk2022audiogen}
Felix Kreuk, Gabriel Synnaeve, Adam Polyak, Uriel Singer, Alexandre D{\'e}fossez, Jade Copet, Devi Parikh, Yaniv Taigman, and Yossi Adi,
\newblock ``Audiogen: Textually guided audio generation,''
\newblock in {\em Proc. ICLR}, 2023.

\bibitem{huang2023make}
Rongjie Huang, Jiawei Huang, Dongchao Yang, Yi~Ren, Luping Liu, Mingze Li, Zhenhui Ye, Jinglin Liu, Xiang Yin, and Zhou Zhao,
\newblock ``Make-an-audio: Text-to-audio generation with prompt-enhanced diffusion models,''
\newblock in {\em Proc. ICML}, 2023.

\bibitem{ghosal2023text}
Deepanway Ghosal, Navonil Majumder, Ambuj Mehrish, and Soujanya Poria,
\newblock ``Text-to-audio generation using instruction-tuned llm and latent diffusion model,''
\newblock in {\em Proc. ACM MM}, 2023.

\bibitem{evans2024fast}
Zach Evans, CJ~Carr, Josiah Taylor, Scott~H Hawley, and Jordi Pons,
\newblock ``Fast timing-conditioned latent audio diffusion,''
\newblock in {\em Proc. ICML}, 2024.

\bibitem{liu2024audiolcm}
Huadai Liu, Rongjie Huang, Yang Liu, Hengyuan Cao, Jialei Wang, Xize Cheng, Siqi Zheng, and Zhou Zhao,
\newblock ``Audiolcm: Text-to-audio generation with latent consistency models,''
\newblock in {\em Proc. ACM MM}, 2024.

\bibitem{xue2024auffusion}
Jinlong Xue, Yayue Deng, Yingming Gao, and Ya~Li,
\newblock ``Auffusion: Leveraging the power of diffusion and large language models for text-to-audio generation,''
\newblock {\em IEEE/ACM Trans. on Audio, Speech, and Language Processing}, 2024.

\bibitem{popov2021grad}
Vadim Popov, Ivan Vovk, Vladimir Gogoryan, Tasnima Sadekova, and Mikhail Kudinov,
\newblock ``Grad-tts: A diffusion probabilistic model for text-to-speech,''
\newblock in {\em Proc. ICML}, 2021.

\bibitem{kim2020glow}
Jaehyeon Kim, Sungwon Kim, Jungil Kong, and Sungroh Yoon,
\newblock ``Glow-tts: A generative flow for text-to-speech via monotonic alignment search,''
\newblock {\em Proc. NeurIPS}, 2020.

\bibitem{kim22guidedtts}
Heeseung Kim, Sungwon Kim, and Sungroh Yoon,
\newblock ``Guided-tts: {A} diffusion model for text-to-speech via classifier guidance,''
\newblock in {\em Proc. ICML}, 2022.

\bibitem{huang2023prosody}
Rongjie Huang, Chunlei Zhang, Yi~Ren, Zhou Zhao, and Dong Yu,
\newblock ``Prosody-tts: Improving prosody with masked autoencoder and conditional diffusion model for expressive text-to-speech,''
\newblock in {\em Findings of the Association for Computational Linguistics: ACL}, 2023.

\bibitem{kim2023crossspeech}
Ji-Hoon Kim, Hong-Sun Yang, Yoon-Cheol Ju, Il-Hwan Kim, and Byeong-Yeol Kim,
\newblock ``Crossspeech: Speaker-independent acoustic representation for cross-lingual speech synthesis,''
\newblock in {\em Proc. ICASSP}, 2023.

\bibitem{nguyen2024fregrad}
Tan~Dat Nguyen, Ji-Hoon Kim, Youngjoon Jang, Jaehun Kim, and Joon~Son Chung,
\newblock ``Fregrad: Lightweight and fast frequency-aware diffusion vocoder,''
\newblock in {\em Proc. ICASSP}, 2024.

\bibitem{Ju24naturalspeech}
Zeqian Ju, Yuancheng Wang, Kai Shen, Xu~Tan, Detai Xin, Dongchao Yang, Eric Liu, Yichong Leng, Kaitao Song, Siliang Tang, et~al.,
\newblock ``Naturalspeech 3: Zero-shot speech synthesis with factorized codec and diffusion models,''
\newblock in {\em Proc. ICML}, 2024.

\bibitem{lee2024ditto}
Keon Lee, Dong~Won Kim, Jaehyeon Kim, and Jaewoong Cho,
\newblock ``Ditto-tts: Efficient and scalable zero-shot text-to-speech with diffusion transformer,''
\newblock {\em arXiv:2406.11427}, 2024.

\bibitem{rombach2022high}
Robin Rombach, Andreas Blattmann, Dominik Lorenz, Patrick Esser, and Bj{\"o}rn Ommer,
\newblock ``High-resolution image synthesis with latent diffusion models,''
\newblock in {\em Proc. CVPR}, 2022.

\bibitem{song2020score}
Yang Song, Jascha Sohl-Dickstein, Diederik~P Kingma, Abhishek Kumar, Stefano Ermon, and Ben Poole,
\newblock ``Score-based generative modeling through stochastic differential equations,''
\newblock in {\em Proc. ICLR}, 2020.

\bibitem{ho2020denoising}
Jonathan Ho, Ajay Jain, and Pieter Abbeel,
\newblock ``Denoising diffusion probabilistic models,''
\newblock {\em Proc. NeurIPS}, 2020.

\bibitem{liu2023audioldm}
Haohe Liu, Zehua Chen, Yi~Yuan, Xinhao Mei, Xubo Liu, Danilo Mandic, Wenwu Wang, and Mark~D Plumbley,
\newblock ``Audioldm: Text-to-audio generation with latent diffusion models,''
\newblock in {\em Proc. ICML}, 2023.

\bibitem{lee2024voiceldm}
Yeonghyeon Lee, Inmo Yeon, Juhan Nam, and Joon~Son Chung,
\newblock ``Voiceldm: Text-to-speech with environmental context,''
\newblock in {\em Proc. ICASSP}, 2024.

\bibitem{vyas2023audiobox}
Apoorv Vyas, Bowen Shi, Matthew Le, Andros Tjandra, Yi-Chiao Wu, Baishan Guo, Jiemin Zhang, Xinyue Zhang, Robert Adkins, William Ngan, et~al.,
\newblock ``Audiobox: Unified audio generation with natural language prompts,''
\newblock {\em arXiv:2312.15821}, 2023.

\bibitem{liu2024audioldm}
Haohe Liu, Yi~Yuan, Xubo Liu, Xinhao Mei, Qiuqiang Kong, Qiao Tian, Yuping Wang, Wenwu Wang, Yuxuan Wang, and Mark~D Plumbley,
\newblock ``Audioldm 2: Learning holistic audio generation with self-supervised pretraining,''
\newblock {\em IEEE/ACM Trans. on Audio, Speech, and Language Processing}, vol. 32, 2024.

\bibitem{yang2023uniaudio}
Dongchao Yang, Jinchuan Tian, Xu~Tan, Rongjie Huang, Songxiang Liu, Xuankai Chang, Jiatong Shi, Sheng Zhao, Jiang Bian, Xixin Wu, et~al.,
\newblock ``Uniaudio: An audio foundation model toward universal audio generation,''
\newblock {\em arXiv:2310.00704}, 2023.

\bibitem{wu2023large}
Yusong Wu, Ke~Chen, Tianyu Zhang, Yuchen Hui, Taylor Berg-Kirkpatrick, and Shlomo Dubnov,
\newblock ``Large-scale contrastive language-audio pretraining with feature fusion and keyword-to-caption augmentation,''
\newblock in {\em Proc. ICASSP}, 2023.

\bibitem{zen2019libritts}
Heiga Zen, Viet Dang, Rob Clark, Yu~Zhang, Ron~J Weiss, Ye~Jia, Zhifeng Chen, and Yonghui Wu,
\newblock ``Libritts: A corpus derived from librispeech for text-to-speech,''
\newblock in {\em Proc. Interspeech}, 2019.

\bibitem{ardila2019common}
Rosana Ardila, Megan Branson, Kelly Davis, Michael Henretty, Michael Kohler, Josh Meyer, Reuben Morais, Lindsay Saunders, Francis~M Tyers, and Gregor Weber,
\newblock ``Common voice: A massively-multilingual speech corpus,''
\newblock {\em arXiv:1912.06670}, 2019.

\bibitem{nagrani2017voxceleb}
Arsha Nagrani, Joon~Son Chung, and Andrew Zisserman,
\newblock ``Voxceleb: a large-scale speaker identification dataset,''
\newblock {\em arXiv:1706.08612}, 2017.

\bibitem{chung2018voxceleb2}
Joon~Son Chung, Arsha Nagrani, and Andrew Zisserman,
\newblock ``Voxceleb2: Deep speaker recognition,''
\newblock in {\em Proc. Interspeech}, 2018.

\bibitem{kwak2024voxmm}
Doyeop Kwak, Jaemin Jung, Kihyun Nam, Youngjoon Jang, Jee-Weon Jung, Shinji Watanabe, and Joon~Son Chung,
\newblock ``Voxmm: Rich transcription of conversations in the wild,''
\newblock in {\em Proc. ICASSP}, 2024.

\bibitem{mei2024wavcaps}
Xinhao Mei, Chutong Meng, Haohe Liu, Qiuqiang Kong, Tom Ko, Chengqi Zhao, Mark~D Plumbley, Yuexian Zou, and Wenwu Wang,
\newblock ``Wavcaps: A chatgpt-assisted weakly-labelled audio captioning dataset for audio-language multimodal research,''
\newblock {\em IEEE/ACM Trans. on Audio, Speech, and Language Processing}, vol. 32, 2024.

\bibitem{koizumi2023libritts}
Yuma Koizumi, Heiga Zen, Shigeki Karita, Yifan Ding, Kohei Yatabe, Nobuyuki Morioka, Michiel Bacchiani, Yu~Zhang, Wei Han, and Ankur Bapna,
\newblock ``Libritts-r: A restored multi-speaker text-to-speech corpus,''
\newblock in {\em Proc. Interspeech}, 2023.

\bibitem{jung2023metric}
Jaemin Jung, Youkyum Kim, Jihwan Park, Youshin Lim, Byeong-Yeol Kim, Youngjoon Jang, and Joon~Son Chung,
\newblock ``Metric learning for user-defined keyword spotting,''
\newblock in {\em Proc. ICASSP}, 2023.

\bibitem{radford2023robust}
Alec Radford, Jong~Wook Kim, Tao Xu, Greg Brockman, Christine McLeavey, and Ilya Sutskever,
\newblock ``Robust speech recognition via large-scale weak supervision,''
\newblock in {\em Proc. ICML}, 2023.

\bibitem{peebles2023scalable}
William Peebles and Saining Xie,
\newblock ``Scalable diffusion models with transformers,''
\newblock in {\em Proc. ICCV}, 2023.

\bibitem{chen2023pixartalpha}
Junsong Chen, Jincheng Yu, Chongjian Ge, Lewei Yao, Enze Xie, Yue Wu, Zhongdao Wang, James Kwok, Ping Luo, Huchuan Lu, et~al.,
\newblock ``Pixart-$\alpha$: Fast training of diffusion transformer for photorealistic text-to-image synthesis,''
\newblock {\em arXiv:2310.00426}, 2023.

\bibitem{beltagy2020longformer}
Iz~Beltagy, Matthew~E Peters, and Arman Cohan,
\newblock ``Longformer: The long-document transformer,''
\newblock {\em arXiv:2004.05150}, 2020.

\bibitem{vaswani2017attention}
A~Vaswani,
\newblock ``Attention is all you need,''
\newblock in {\em Proc. NeurIPS}, 2017.

\bibitem{perez2018film}
Ethan Perez, Florian Strub, Harm De~Vries, Vincent Dumoulin, and Aaron Courville,
\newblock ``Film: Visual reasoning with a general conditioning layer,''
\newblock in {\em Proc. AAAI}, 2018.

\bibitem{wang2024v2a}
Heng Wang, Jianbo Ma, Santiago Pascual, Richard Cartwright, and Weidong Cai,
\newblock ``V2a-mapper: A lightweight solution for vision-to-audio generation by connecting foundation models,''
\newblock in {\em Proc. AAAI}, 2024.

\bibitem{radford2021learning}
Alec Radford, Jong~Wook Kim, Chris Hallacy, Aditya Ramesh, Gabriel Goh, Sandhini Agarwal, Girish Sastry, Amanda Askell, Pamela Mishkin, Jack Clark, et~al.,
\newblock ``Learning transferable visual models from natural language supervision,''
\newblock in {\em International conference on machine learning}. PMLR, 2021.

\bibitem{kong2020hifi}
Jungil Kong, Jaehyeon Kim, and Jaekyoung Bae,
\newblock ``Hifi-gan: Generative adversarial networks for efficient and high fidelity speech synthesis,''
\newblock in {\em Proc. NeurIPS}, 2020.

\bibitem{chen2022wavlm}
Sanyuan Chen, Chengyi Wang, Zhengyang Chen, Yu~Wu, Shujie Liu, Zhuo Chen, Jinyu Li, Naoyuki Kanda, Takuya Yoshioka, Xiong Xiao, et~al.,
\newblock ``Wavlm: Large-scale self-supervised pre-training for full stack speech processing,''
\newblock {\em IEEE Journal of Selected Topics in Signal Processing}, vol. 16, 2022.

\bibitem{jung2024espnet}
Jee-weon Jung, Wangyou Zhang, Jiatong Shi, Zakaria Aldeneh, Takuya Higuchi, Barry-John Theobald, Ahmed~Hussen Abdelaziz, and Shinji Watanabe,
\newblock ``Espnet-spk: full pipeline speaker embedding toolkit with reproducible recipes, self-supervised front-ends, and off-the-shelf models,''
\newblock in {\em Proc. Interspeech}, 2024.

\bibitem{loshchilov2017decoupled}
Ilya Loshchilov and Frank Hutter,
\newblock ``Decoupled weight decay regularization,''
\newblock in {\em Proc. ICLR}, 2017.

\bibitem{kilgour2018fr}
Kevin Kilgour, Mauricio Zuluaga, Dominik Roblek, and Matthew Sharifi,
\newblock ``Fr$\backslash$'echet audio distance: A metric for evaluating music enhancement algorithms,''
\newblock {\em arXiv:1812.08466}, 2018.

\bibitem{sheffer2023hear}
Roy Sheffer and Yossi Adi,
\newblock ``I hear your true colors: Image guided audio generation,''
\newblock in {\em Proc. ICASSP}, 2023.

\bibitem{iashin2021taming}
Vladimir Iashin and Esa Rahtu,
\newblock ``Taming visually guided sound generation,''
\newblock in {\em Proc. BMVC.}, 2021.

\end{thebibliography}

\end{document}